\documentclass[preprint,preprintnumbers,amsmath,amssymb]{revtex4}

\usepackage{graphicx}
\usepackage{dcolumn}
\usepackage{bm}
\begin{document}

\title{Persistent Current in Two Coupled Rings}

\author{D. Schmeltzer and Hsuan-Yeh Chang}

\affiliation{Physics Department, City College of the City University of New York \\
New York, New York 10031}

\date{\today}
\begin{abstract}
We report the solution of the persistent current in two coupled rings in the presence of external magnetic fluxes. We showed that the magnetic fluxes  modify the global phase  of the electronic wave function for  multiple connected geometry formed by the coupled rings.  We obtained an exact solution for the persistent current and investigated the exact solution numerically.  For two large coupled rings with equal fluxes, we found that the persistent current in the two coupled rings is in fact equal to that in a single ring.  This theory explains the experimental results observed in a line of sixteen coupled rings.  (Phys. Rev. Lett. 86, 3124 (2001).)  
\end{abstract}

\maketitle
\textbf{Introduction.}
In Quantum Mechanics, the wave function is path dependent and is sensitive to the presence of a vector potential caused by  an external magnetic flux.  In a closed geometrical structure, such as a ring, the wave function is changed by a measurable phase [1], causing all the physical properties to become periodic [2].  When a mesoscopic ring of normal metal is pierced by a magnetic flux $\Phi$ [2], the boundary conditions are modified, leading to a famous theorem of periodic properties with the flux period $\Phi_0 = h / e$ and to a remarkable phenomena [3] of a non-dissipative \textit{persistent current} [3-5]. 

One way to classify the closed geometrical structure is by using the number of $holes$ formed on the closed geometrical structure.  For a $closed$ $surface$, the number of holes formed thereon is often referred to as a genus number $g$ [6, 9].  For example, a genus number $g=1$ describes an Aharonov-Bohm ring geometry, while a genus number $g = 2$ describes two rings perfectly glued at one point to form a character ``8'' structure. 

In spite of the predictions that the global geometry [1] has to affect the wave function, persistent currents were believed to be a specific property of isolated single rings. Only a recent experiment of 16 $GaAs/GaAlAs$ connected rings [12] has demonstrated the possibility of $persistent$ current in connected rings, namely, in a material with a large genus number.  The experimental result showed that the current in a 16-rings structure is approximately the same as the current in a single ring, which disagrees with the existing  theoretical results [11].

In this Letter, we report that for multiple connected geometries, such as a geometry with two or more holes (or connected rings), the geometry modifies the $global$ properties of the wave function, and the presence of magnetic fluxes generates persistent currents with complicated periods.  We present an exact analytical solution for the eigenvalues and  compute the persistent current for two coupled rings with a character ``8'' structure for two different fluxes.  We solve the problem by modeling the $gluing$ of the two rings using $fermionic$ constraints [14] with $anti$-$commuting$ $Lagrange$ multipliers, which can be viewed as a resonant impurity strongly coupled to the two rings [13]. 
  
The analytical results are investigated numerically.  When the two fluxes on both rings are the same, we find a simple relation between the single ring ($g=1$) current, $I^{(g=1)}(flux;N)$, and the double ring ($g=2$) current, $I^{(g=2)}(flux;N)$.  At $T=0$, we define $ I^{(g=2)}(flux;N) = r(N) I^{(g=1)}(flux;N)$, where $r(N)$ is a ratio between the two currents.  The ratio $r(N)$ is a function of the number of sites $N$ and obeys $r(N) \rightarrow 1$ for $N \rightarrow \infty$.  This result explains the experimental puzzle for a line of 16 GaAs/GaAlAs connected rings [12].

\textbf{Exact Solution for Two Coupled Rings.}
We consider the Hamiltonian $H_{0}$ for two spinless fermionic rings in the $absence$ of a magnetic flux. The rings obey periodic boundary conditions. For each ring, the point $x$ is identified with the point $x+L$.  The two coupled rings with the character ``8'' structure (i.e. $g=2$) are obtained by identifying the middle point $x = L/2$ of the first ring with point $x=0$ of the second ring, i.e. $C_1(L/2) = C_2(0)$ and $C^+_1(L/2) = C^+_2(0)$.  This identification is equivalent to two $fermionic$ $constraints$, $Q \equiv C_1(L/2) - C_2(0)$ and $Q^+ \equiv C^+_1(L/2) - C^+_2(0)$.  Since the constraints are fermionic, they can be enforced by using $anti$-$commuting$ Lagrange multipliers, $\mu^+$ and $\mu$.  Following ref. [14], we introduce the Hamiltonian with the constraints,
\[
H_{T} = H + \mu^+ Q + Q^+ \mu
\]

The  unusual physical meaning of the anti-commuting Lagrange multipliers can be viewed as a fermionic impurity [13], which mediates the hopping of the electrons between the two rings. 
The two rings of length $L$ are threaded by a magnetic flux $\Phi_\alpha$, where $\alpha = 1, 2$ (for each ring).  In order to observe the changes of the constraints in the presence of the external flux, we perform the following steps. In the $absence$ of the $external$ $flux$ $\Phi_\alpha$, the annihilation and creation fermion operators obey periodic boundary conditions $C_\alpha(x) = C_\alpha (x + L)$ and $C^+_\alpha(x) = C^+_\alpha (x + L)$, where $\alpha = 1, 2$. The genus $g=2$ is implemented by the Fermionic constraints $Q = C_1(L/2) - C_2(0)$ and $Q^+ = C^+_1(L/2) - C^+_2(0)$, and Hamiltonian $H_0 = - t \sum_{\alpha=1}^2 \sum_{x = 0}^{(N_{s} - 1) a} [C_\alpha^+(x) C_\alpha(x + a) + h.c.]$. The length of each ring is $L = N_{s} a$, where $N_{s}$ is the number of sites and $a$ is the lattice spacing. When  the  $external$ $magnetic$ $flux$ $\Phi_\alpha$ is applied  the Hamiltonian $H_0$ is replaced by $H$. The Hamiltonian $H$ is obtained
by the transformation $C_\alpha(x) \rightarrow \exp[i \frac{e}{\hbar c} \int_0^x A (x') d x' ] C_\alpha(x) \equiv \psi_\alpha(x)$ and $C^+_\alpha(x) \rightarrow C^+_\alpha(x) \exp[ - i \frac{e}{\hbar c} \int_0^x A (x') d x' ] \equiv \psi^+_\alpha(x)$. Here $A(x)$ is the $tangential$ component of vector potential on the ring. The relation between the flux and the vector potential on each ring is $\frac{e}{\hbar c} \int_0^L A(x) dx \simeq \frac{e}{\hbar c} A_{\alpha} N_{s} a$.

The flux $\Phi_\alpha$ on each ring $\alpha=1, 2$ gives rise to a change in the boundary conditions, $\psi_\alpha(x + N_{s} a)= \psi_\alpha(x) e^{i \varphi_\alpha}$ and $\psi^+_\alpha(x + N_{s} a) = \psi^+_\alpha(x) e^{- i \varphi_\alpha}$, where $\varphi_\alpha = 2 \pi (\frac{e \Phi_\alpha}{h c} ) = 2 \pi \frac{\Phi_\alpha}{\Phi_0} \equiv 2 \pi \hat{\varphi}_\alpha$. This boundary condition gives rise to a normal mode expression for each ring, $\psi_\alpha(x) = \frac{1}{\sqrt{N}} \sum_{n=0}^{N_{s}-1} e^{i K(n, \varphi_\alpha) \cdot x} \psi_\alpha(n)$ and a similar expression for $\psi^+_\alpha(x)$. The ``momentum'' $K(n, \varphi_\alpha)$ is given by
\[
K(n, \varphi_\alpha ) = \frac{2 \pi}{N_{s} a} (n + \hat{\varphi}_\alpha)
\]
where $n = 0, 1, \ldots, N-1$ are integers with $N = N_{s}$, and $\varphi_\alpha = 2 \pi \hat{\varphi}_\alpha$. In the momentum space, the Fermionic operators $\psi_\alpha(n)$ and $\psi^+_\beta(m)$ obey anti-commutation relations, $[ \psi_\alpha(n), \psi^+_\beta (m) ]_+ = \delta_{\alpha, \beta} \delta_{n,m}$. The Hamiltonian for the two rings in the presence the external flux takes the form,
\begin{equation} \label{eq:hamiltonian}
H = - t \sum_{\alpha = 1, 2} \sum_{x = 0}^{(N_{s}-1) a} \psi^+_\alpha (x) \psi_\alpha(x + a) + h.c. = \sum_{\alpha = 1, 2} \sum_{n = 0}^{N_{s}-1} \epsilon(n, \hat{\varphi}_\alpha) \psi^+_\alpha(n) \psi_\alpha (n)
\end{equation}
where $\epsilon(n, \varphi_\alpha) = - 2 t \cos [ \frac{2 \pi}{N} (n + \hat{\varphi}_\alpha)]$ are the eigenvalues for each ring. The Hamiltonian in eq. \ref{eq:hamiltonian} has to be solved $together$ with the $transformed$ $constraints$ equations,$Q = \psi_1(\frac{L}{2}) e^{- i \varphi_1} - \psi_2(0)$ and $Q^+ = \psi^+_1(\frac{L}{2}) e^{ i \varphi_1} - \psi^+_2(0)$.

The wave function for the genus $g=2$ problem is given by the eigenstate $ | \chi \rangle $ of the Hamiltonian in eq. \ref{eq:hamiltonian}, which in addition satisfies the equations $Q | \chi \rangle = 0$ and $Q^+ | \chi \rangle = 0$. The constraint conditions are implemented with the help of the $anti-commuting$ $Lagrange$ $multipliers$ $\mu$ and $\mu^+$. The Hamiltonian $H_{T}$ with the constraints takes the form,
\begin{equation}
H_{T} = H + \mu^+ Q + Q^+ \mu.
\end{equation}
The Lagrange multiplier are determined by the condition that the constraints are satisfied at any time. Therefore, the time derivative satisfies the equation, $\dot{Q} | \chi \rangle = \dot{Q}^+ | \chi \rangle = 0$ at any time. We will use the notations, $[A, B]_+ \equiv A B + B A$ and $ [A, B] = A B - B A$.
The Heisenberg equation of motion for the constraint $Q$ is,
\begin{eqnarray} \nonumber
i \hbar \dot{Q}&=& [Q, H_{T}] = [Q, H] + [Q, \mu^+ Q + Q^+ \mu ] \\ \nonumber
               &=& [Q, H] + [Q, \mu^+]_+ Q - \mu^+ [ Q, Q]_+ + [Q, Q^+]_+ \mu - Q^+ [ Q, \mu]_+ \\ \label{eq:heom}
               &=& [Q, H] + [Q, Q^+]_+ \mu
\end{eqnarray}
The rest of the anti-commutators in eq. \ref{eq:heom} vanishes. The anti-commuting Lagrange multipliers satisfy, $[Q, \mu^+]_+ = [Q, \mu]_+ = [Q^+, \mu^+]_+ = [Q^+, \mu]_+ = 0$. Since the constraints are fermionic, we obtain that they obey $[Q, Q^+]_+ = [Q^+, Q]_+ = 2$. Therefore, the constraints are second class constraints [14]. From the condition $\dot{Q} | \chi \rangle = 0$ and eq. \ref{eq:heom}, we determine the Lagrange multiplier field $\mu$.
\begin{equation}
\mu = - [Q^+, Q]_+^{-1} [Q, H] = - \frac{1}{2} [Q, H]
\label{eq:mu}
\end{equation}
$\mu^+$ is obtained from the  equation  $\dot{Q}^+| \chi \rangle = 0$,
\begin{equation}
\mu^+ = [ Q, Q^+]_+^{-1} [Q^+, H] = \frac{1}{2} [Q^+, H]
\label{eq:mustar}
\end{equation}

The Hamiltonian $H_{T}$ with the constraints and the $Lagrange$ multipliers are used to compute the $Heisenberg$ $equation$ of $motion$ for any $Fermionic$ $operator$, $\hat{O}$. (The Lagrange multipliers anti-commute with any Fermionic operator, i.e. $[\hat{O}, \mu]_+ = [\hat{O}, \mu^+]_+ = 0$.)
\begin{eqnarray} \nonumber
i \hbar \frac{d \hat{O}}{d t} = [ \hat{O}, H_{T}] &=& [\hat{O}, H] + [\hat{O}, \mu^+ Q] + [ \hat{O}, Q^+] \mu \\ \nonumber
&=& [\hat{O}, H] + [\hat{O}, \mu^+ ]_+ Q - \mu^+ [ \hat{O}, Q]_+ + [\hat{O}, Q^+]_+ \mu - Q^+ [ \hat{O}, \mu]_+ \\ \label{eq:oeom}
&=& [\hat{O}, H] - [\hat{O}, Q]_+ \mu^+ - [\hat{O}, Q^+] \mu
\end{eqnarray}
We substitute in eq. \ref{eq:oeom} the solution for the Lagrange multiplier fields given by eqs. \ref{eq:mu} and \ref{eq:mustar}. We obtain a $new$ $equation$ of $motion$ with a $new$ $commutator$, which resemble the classical Dirac brackets [13].
\begin{eqnarray} \nonumber
i \hbar \frac{d \hat{O}}{d t} = [ \hat{O}, H_{T}] &=& [\hat{O}, H] - [\hat{O}, Q^+]_+ ([Q^+, Q]_+)^{-1} [Q, H] - [\hat{O}, Q]_+ ([Q, Q^+]_+)^{-1} [Q^+, H] \\ \label{eq:dirac}
&\equiv& [\hat{O}, H ]_D
\end{eqnarray}
Eq. \ref{eq:dirac} shows that the Heisenberg equation of motion is governed by a $new$ $commutator$, $[\hat{O}, H ]_D$. We will use this new commutator to compute the  Heisenberg equations of motion for the creation and annihilation Fermionic operators $\psi_\alpha(x,t)$ and $\psi_\alpha^+(x,t)$, where $\alpha = 1, 2$.
\begin{eqnarray} \nonumber
i \hbar \dot{\psi}_\alpha(x) &=& [ \psi_\alpha(x), H]_D = [ \psi_\alpha(x), H] - \frac{1}{2}[ \psi_\alpha(x), Q^+]_+ [Q,H] \\ \nonumber
&=& - t [ \psi_\alpha(x + a)+ \psi_\alpha(x-a)] - \frac{1}{2}[ \delta_{\alpha, 1} \delta_{x, L/2} e^{i \varphi_1} - \delta_{\alpha, 2} \delta_{x, 0}] \\ \label{eq:4}
& & \cdot (-t) \{ e^{-i\varphi_1} [ \psi_1(\frac{L}{2} + a) + \psi_1(\frac{L}{2}-a) ] +
e^{-i\varphi_2} [ \psi_2(\frac{L}{2} + a) + \psi_2(\frac{L}{2}-a) ] \}
\end{eqnarray}

The ground state wave function is obtained from the one electron state, $|\chi> = \sum_{\alpha = 1, 2} \sum_{x = 0}^{(N_{s}-1) a} Z_\alpha(x) \psi_{\alpha}^+ (x) |0 >$, given in terms of the site amplitudes $Z_\alpha(x)$. Using a normal mode momentum expansion, $f_\alpha(n)$, i.e. $Z_\alpha(x) = \frac{1}{\sqrt{N}}\sum_{n=0}^{N-1} e^{i K(n, \hat{\varphi}_\alpha) x} f_{\alpha}(n)$, we find the following equations for the eigenvalues $\lambda$ and the amplitudes in the momentum space $f_{\alpha}(n)$,
\begin{equation} \label{eq:normal1}
(\lambda - \epsilon(\ell + \hat{\varphi}_1)) f_1(\ell) = - \frac{e^{i \pi \ell}}{2N} \sum_{n=0}^{N-1} \epsilon(n + \hat{\varphi}_1) e^{i \pi n} f_1(n) - \frac{1}{2N} \sum_{n=0}^{N-1} \epsilon(n + \hat{\varphi}_2) f_2(n)
\end{equation}
and
\begin{equation} \label{eq:normal2}
(\lambda - \epsilon(\ell + \hat{\varphi}_2)) f_2(\ell) = \frac{1}{2N} \sum_{n=0}^{N-1} \epsilon(n + \hat{\varphi}_2) f_2(n) + \frac{e^{i \pi \ell}}{2N} \sum_{n=0}^{N-1} \epsilon(n + \hat{\varphi}_1)e^{i \pi n} f_1(n).
\end{equation}

We diagonalize eqs. \ref{eq:normal1} and \ref{eq:normal2} by linear transformations,
$S_1 (\hat{\varphi}_1, \lambda) = -\sum_{\ell = 0}^{N-1} \epsilon(\ell+\hat{\varphi}_1) e^{i\pi\ell} f_1(\ell)$ and $S_2 (\hat{\varphi}_2, \lambda) = -\sum_{\ell = 0}^{N-1} \epsilon(\ell+\hat{\varphi}_2) f_2(\ell)$.
As a result, we obtain the equation,
$\textbf{M} \left(%
\begin{array}{c}
  S_1 \\
  S_2 \\
\end{array}
\right) = 0$,
where the matrix $\textbf{M}$ is given by $\textbf{M} = \left(%
\begin{array}{cc}
  -(1+\Delta^{(+)}_1) & \Delta^{(-)}_1 \\
       \Delta^{(-)}_2 & 1 - \Delta^{(+)}_2 \\
\end{array}%
\right)$.
Here, we define $\Delta^{(+)}_{\alpha}(\hat{\varphi}_\alpha, \lambda )\equiv\Delta_{\alpha}^{(even)} (\hat{\varphi}_\alpha, \lambda) + \Delta_{\alpha}^{(odd)} (\hat{\varphi}_\alpha, \lambda)$ and $\Delta^{(-)}_{\alpha}(\hat{\varphi}_\alpha, \lambda )\equiv\Delta_{\alpha}^{(even)} (\hat{\varphi}_\alpha, \lambda) - \Delta_{\alpha}^{(odd)} (\hat{\varphi}_\alpha, \lambda)$, with the $even$ and $odd$ representations given by, $\Delta_{\alpha}^{(even)} (\hat{\varphi}_\alpha, \lambda)=\frac{1}{2N} \sum_{m = 0}^{(N-2)/2} \frac{\epsilon( 2m + \hat{\varphi}_\alpha ) }{ \lambda - \epsilon(2 m + \hat{\varphi}_\alpha)}$ and $\Delta_{\alpha}^{(odd)} (\hat{\varphi}_\alpha, \lambda)=\frac{1}{2N} \sum_{m = 0}^{(N-2)/2} \frac{\epsilon( 2m + 1 + \hat{\varphi}_\alpha ) }{ \lambda - \epsilon(2 m + 1 + \hat{\varphi}_\alpha)}$. We compute $\det \textbf{M} = 0$ and obtain the $characteristic$ $polynomial$ which is used to determine  the $eigenvalues$ $\lambda$.
\begin{equation}
2 [ \Delta_1^{(even)} (\hat{\varphi}_1, \lambda) \Delta_2^{(odd)} (\hat{\varphi}_2, \lambda) + \Delta_1^{(odd)} (\hat{\varphi}_1, \lambda) \Delta_2^{(even)} (\hat{\varphi}_2, \lambda)] + [\Delta^{(+)}_1 (\hat{\varphi}_1, \lambda) - \Delta^{(+)}_2 (\hat{\varphi}_2, \lambda)] = 1
\label{eq:secular}
\end{equation}
Eq. \ref{eq:secular} is our main result for the genus $g=2$ case. We observe that the matrix $M$ is $symmetric$ and the eigenvalues are real when the fluxes are equal, i.e. $\hat{\varphi}_1=\hat{\varphi}_2$, or opposite, i.e. $\hat{\varphi}_1=-\hat{\varphi}_2$. For other cases, the eigenvalues can have imaginary parts, thereby giving rise to non-conducting states.

\textbf{Numerical Solution and Comparison with Experiment.}
We have numerically solved the secular equation \ref{eq:secular}. To compute the current, we sum over the current carried by each eigenvalue $\lambda(\hat{\varphi}_1,\hat{\varphi}_2)$ using the grand-canonical ensemble. The current in each ring $\alpha=1, 2$ is given by,
\begin{equation}
I^{(g=2)}_{\alpha}(\hat{\varphi}_1,\hat{\varphi}_2) = -\sum_{\lambda(\hat{\varphi}_1,\hat{\varphi}_2)}{\frac{d}{d\hat{\varphi}_\alpha}[ \lambda(\hat{\varphi}_1,\hat{\varphi}_2)}]F(\frac{(\lambda(\hat{\varphi}_1,\hat{\varphi}_2)-E_{fermi})}{K_{Boltzman}T})
\label{eq:current}
\end{equation}
where $F(\frac{(\lambda(\hat{\varphi}_1,\hat{\varphi}_2)-E_{fermi})}{K_{Boltzman}T})\equiv [1+e^{\frac{(\lambda(\hat{\varphi}_1,\hat{\varphi}_2)-E_{fermi})}{K_{Boltzman}T}} ]^{-1}$ is the Fermi Dirac function  which depends on the chemical potential $E_{fermi}$ and temperature $T$. The current is sensitive to the number of electrons being either even or odd. We use the grand-canonical ensemble and limit ourselves to a situation with even numbers of sites and a zero chemical potential, i.e. $E_{fermi}=0$ (which corresponds to the half-filled case). In order to have a perfect particle-hole symmetry, we will $restrict$ the analysis to the $special$ series for the $number$ of $sites$ being $N_{s}= 2, 6, 10, 14, 18, \ldots, 2m+2$, where $m = 0, 1, 2, 3 \ldots$. For this case, we find that, when the fluxes are the same in both rings, the current for $g=2$ has the same periodicity as that of a single ring, i.e. $I^{(g=2)}(\Phi + \Phi_0) = I^{(g=2)}(\Phi)$. At temperatures $T \leq 0.02$ Kelvin, the line shape of the current as a function of the flux is of a $sawtooth$ form (see figure $1b$). For other series $N_{s}\neq 2m+2$, the periodicity of the current is complicated. Using the experimental values given in the experiment [12], we estimate that the number of sites in our model should be in the range of $N_{s}= 50 \sim 150$, the $hopping$ constant should be $t= \frac{\hbar v_{fermi}}{2 a sin(K_{fermi}a)} \approx 0.01 $ eV, and the temperature in the experiment should be $T=0.02$ Kelvin. Using these units, we obtain that the persistent current is given in terms of a $dimensionless$ $current$, $I_{\alpha}$ (see figure $1b$ and figure $1c$) with the actual current value, $I^{(g=2)}_{\alpha}= I_{\alpha}\times 2.5\times 10^{-3}$ Ampere.

a) $Equal$ $fluxes$ $for$ $g=2$: For this case the secular equation is simplified and takes the form of $ 4 [\Delta^{(even)} (\hat{\varphi}, \lambda) \Delta^{(odd)} (\hat{\varphi}, \lambda)]=1$.

For $N_{s}=2$, we solve analytically the secular equation. We find that the eigenvalues are given by $\lambda(n,\varphi;N=2)=r(N=2)\epsilon(n,\varphi;N=2)$, where $\epsilon(n, \varphi,N=2) = - 2 t \cos [ \frac{2 \pi}{N=2} (n + \hat{\varphi})]$, and $n=0, 1$ are the single ring eigenvalues. The value for $r(N=2)$ is $r(N=2)=\frac{\sqrt{3}}{2}$. To find the eigenvalues for other number of sites, $N_{s}= 6, 10, 14, 18, 22, 26, 30$, we numerically find the relation, $\lambda(n,\varphi;N)=r(N)\epsilon(n, \varphi;N)$, where $n = 0, 1, \ldots, N-1$ and  $\epsilon(n, \varphi;N)= - 2 t \cos [ \frac{2 \pi}{N} (n + \hat{\varphi})]$ are the single ring eigenvalues.  The function $r(N)$ is given in $figure$ $1a$. This figure shows that the function $r(N)$ reaches $one$ for large $N$. Using the function $r(N)$ given in figure $1a$, we compute the current for the $g=2$ case as a function of temperature.
\begin{equation}
I^{(g=2)}(\varphi;N;T)=-\sum_{n=0}^{n=N-1}\frac{d} {d\varphi}[r(N)\cdot\epsilon(n, \varphi;N)]F(\frac{r(N)\cdot\epsilon(n, \varphi;N)-E_{fermi})}{K_{Boltzman}T})
\label{gcurrent}
\end{equation}
 
$Figure$ $1b$ represents the current computed from eqs. \ref{eq:secular} and \ref{eq:current} for $N_{s}=30$ sites at two temperatures $T=0.02$ and $T=20$ $Kelvin$. In this figure, the current is given in dimensionless units $I$ plotted as a function of the dimensionless flux $\hat{\varphi}_\alpha=[-0.5,0.5]$ ($\varphi_\alpha = 2 \pi \hat{\varphi}_\alpha = [-\pi,\pi]$). The solid line represents the single ring current and the dashed line represents the current for the genus $g=2$ case. In figure $1b$, the ratio of the currents at $T=0.02$ $Kelvin$ is $r(N=30,T=0.02)=0.979$.

The experiment [12] was performed with 16 rings, using the condition that each ring has $50$ $sites$, which gives an estimation of $r = I_{16-rings} / I_{single-ring} \approx [I^{(g=2)} / I_{single-ring}]^{4} = [r(T=0.02,N_{s}=50)]^{4} = [0.987]^{4} = 0.95$. The ratio of  $r=0.95$, is in the range of the experimental observation reported in ref.[12]. 
   
Next, we turn our attention to the values of the currents. The current reported in ref. 12 is in the range of $0.5$ $nA$. Our estimate for 30 sites given in figure $1b$ at $T=0.02$ is $I = 10^{-3}$, which corresponds to a current of $I^{(g=2)}= I \times 2.5 \times 10^{-3}A = 2.5 \times 10^{3}$ $nA$. This means that our current is $10^{3}$ larger then the reported current [12]. The origin of this discrepancy might be due to the following factors: the effective number of sites is larger then 30; the temperature in the experiment might be higher then $T=0.02$ $Kelvin$; and the $elastic$ scattering length $l_{elastic}$ in the experiment is smaller than the length of the ring $L$. Assuming a strong $2K_{F}$ impurity scattering in the presence of the electron-electron interaction can cause a significant current suppression [7,8]. For an impurity with a  transmission coefficient $\hat{t}<1$  and a repulsive electron-electron  interaction  with a $Luttinger$ parameter $K_{c}<1$  the reduction of the current is $(|\hat{t}|^{2})(\frac{\lambda_{fermi}}{L})^{2(K_{c}^{-1}-1)}$ ,[7,8]. The ratio $(\frac{\lambda_{fermi}}{L})\approx\frac{1}{150}$ with $K_{c}=0.6$ gives a suppression of $1000$ for the current, in agreement with the observed currents.

At $T = 20$ Kelvin, the value of the currents are in the range of $7$ $nA$ and the reduction of the current is larger in comparison with the $T = 0.02$ Kevin.

We have checked the behavior of the current as a function of the $number$ $of$ $sites$ (for the single ring and the double ring) at a fixed flux $0.001$ and different temperatures.  At $T = 0.02$ Kelvin the current scales like $\frac{1}{N{s}}$  (for the double ring and the single ring) up to  $N{s}\approx 60$ . Increasing the temperature to $T=1.$ Kelvin the current decreases faster than $\frac{1}{N{s}}$ for $N{s}>30$ . This results confirm that at low temperature the $Persistent$ current in a double ring decreases linearly with the length.
 
b) $Two$ $coupled$ $rings$ $with$ $opposite$ $fluxes$, i.e. $\hat{\varphi} = \hat{\varphi}_1 = - \hat{\varphi}_2$:

For $N_{s}=2$, the eigenvalues are the $same$ as the one obtained for the same flux case. For $N_{s}= 6, 10, 14, \ldots, 2m+2$, we solve the secular equation given in eq. \ref{eq:secular} and compute the eigenvalues. In $figure$ $2b$, we plot the $total$ $energy$ as a function of the opposite fluxes at $T=0.02K$ for $30$ sites, $E^{(g=2)}(-\hat{\varphi},\hat{\varphi}, N_{s} = 30, T=0.02, K) =  \sum_{n=0}^{n=N-1} [\lambda(-\hat{\varphi}, \hat{\varphi}) F(\frac{(\lambda(-\hat{\varphi},\hat{\varphi})-E_{fermi})}{K_{Boltzman}T})]$ . We observe that the total energy has a $chaotic$ structure due to the strong Backscattering caused by the $common$ point of the two rings. For comparison, we also show in $figure$ $2a$ the total energy for $equal$ fluxes. The energy is parabolic for small fluxes and the current is proportional to the flux. For large values of flux the total energy $E^{(g=2)}$ is a periodic function with the period $\Phi_{0}$.

c) We have also considered the case with $unequal$ fluxes. In $ring$ $one$, the flux was fixed at the values $\hat{\varphi}_1 = 0.1, 0.2, 0.3, 0.4$, and in $ring$ two, the flux $\hat{\varphi}_2$ changes continuous from $-0.05$ to $0.05$. For this case, the current in $ring$ $two$ is linear in the flux $\hat{\varphi}_2$ and is independent on the flux in $ring$ $one$, $\hat{\varphi}_1$.

d) $An$ $effective$ $four$ $terminal$ $circuit$ $model$, which explains the physics of the $g=2$ coupled rings.

The persistent current problem is mapped into a closed circuit with a voltage source $V$. We use the relation between the flux and the voltage [7], $\frac{e V}{h}=\frac{2\pi}{N_{s}a}v_{fermi}(\frac{\Phi}{\Phi_0})$. From this relation, we find that a perfect ring is equivalent to one conducting channel connected to a battery with a voltage $V$. This gives a current $I = \frac{e^{2}}{h} V$. The $g=2$ case is described by two circuits coupled at $x=L/2$. The first ring with a flux is equivalent to a circuit that start at $x=0$ runs to the common  point $x=L/2$ and back to $x=L$. We attach a voltage source $V_{1}=(\frac{e }{h})^{-1}\frac{2\pi}{N_{s}a}v_{fermi}(\frac{\Phi_{1}}{\Phi_0})$ between the points $x=0$ and $x=L$. We do the same for the second ring, where we attach a voltage $V_{2} = - (\frac{e}{h})^{-1} \frac{2\pi}{N_{s}a}v_{fermi} (\frac{\Phi_{2}}{\Phi_0})$.

At the coupling region, we split the point $x=L/2$ into two points $O$ and $O'$, such that a $high$ $resistance$ $R_{\infty}$ is attached between the points $O$ and $O'$. Now, we consider separately the equal and opposite fluxes. For $equal$ $fluxes$, we have to use two opposite voltages $V_{1} = -V_{2} = V/2$. For this case, the current does not pass trough the high resistance region between the points $O$ and $O'$. Instead, it forms a loop with $twice$ the $length$ of the single ring with a $total$ $doubled$ $flux$ and an effective voltage $V$. Therefore, the current is the same as in a single ring, i.e. $I=\frac{e^{2}}{h}V$.

Next, we consider the situation for $opposite$ $fluxes$. In this case, the two batteries obey $V_{1} = V_{2} = V/2$. As a result, the current has to flow through the high resistance region $O$ to $O'$. The currents are opposite in each ring and their value is determined by the $high$ $resistance$ $R_{\infty}$. We find, $I = \frac{1}{2R_{\infty} + (\frac{e^{2}}{h})^{-1}} \approx 0$.

\textbf{Summary.}
In this Letter, we have introduced a method which solves the problem of the $global$  phase of the wave function for geometrical structures with holes, i.e. high genus materials.  This method is applicable to a variety of mesoscopic systems where coherency of wave function is important.

We have found an exact solution for the persistent current in two coupled rings.  By numerical calculations, we have computed the current dependence on the flux, temperature, and the number of sites.

This theory resolved the experimental puzzle [12] that the persistent current in many coupled rings is the same as that of the single ring.

\textbf{Acknowledgment.}
D.S. would like to express particular thanks to Dr. Avadh Saxena for early discussions on the experiments related to this work.  D.S. would also like to thank the Theoretical Division T-11 of Los Alamos National Laboratory and the PSC-CUNY for financial supports.

\begin{bibliography}{99}
1. Y. Aharonov and D.Bohm, Phys. Rev. 115, 485 (1959). \\
2. N. Byers and C.N. Yang, Phys. Rev. Lett. 7, 46 (1961). \\
3. M. Buttiker, Y. Imry, and R. Landauer, Phys. Lett. A. 96, 365 (1983).\\
4. L.P. Levy, G. Dolan, J. Dunsmuir, and H. Bouchiat, Phys. Rev. Lett. 64, 2074 (1990).\\
5. Y. Gefen, Y. Imry, and M.Y. Azbel, Phys. Rev. Lett. 52, 129 (1984). \\
6. H. Aoki, J. Phys. C. 18, 1885-1890 (1981).\\
7. D. Schmeltzer, Phys. Rev. B. 63, 125332 (2001); D. Schmeltzer and R. Berkovits, Physics Letters A 253, 341 (1999).\\
8. C.L. Kane and M.P.A. Fisher, Phys. Rev. Lett. 68, 1220 (1992).\\
9. Ken-Ichi Sasaki and Yoshiyuki Kawazoe, Cond-Mat/0408505.\\
10. K. Sasaki, Y. Kewazoe, and R. Saito, Physics Letters A321, 369-375 (2004).\\
11. M. Pascaud and G. Montambaux, Phys. Rev. Lett. 82, 4512 (1999).\\
12. W. Rabaud, L. Saminadayar, D. Mailly, K. Hasselbach, A. Benoit, and B. Etienne, Phys. Rev. Lett. 86, 3124 (2001).\\
13. P. Mehta and N.Andrei, Phys. Rev. Lett. 96, 216802 (2006).\\
14. Paul A. M. Dirac, ``Lectures on Quantum Mechanics,'' Belfer Graduate School of Science, Yeshiva University, New York, 1964.
\end{bibliography}

\pagebreak

\begin{figure}
\includegraphics[width=4in]{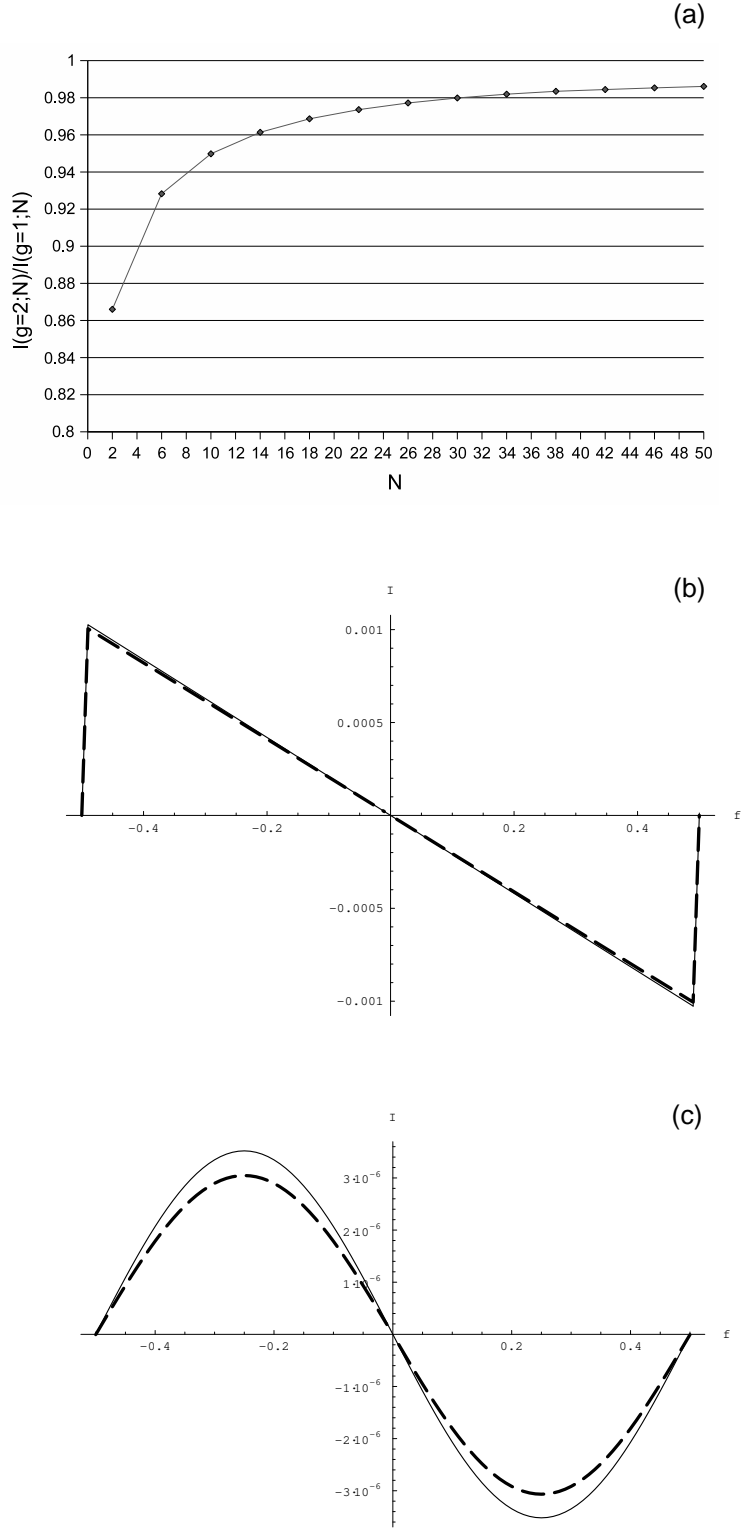}
\caption{(a) The ratio of the double to single ring currents $I(g=2;N)/I(g=1;N) = r(N)$; (b) The single ring (solid line) and the double ring (dashed line) currents for $N_{s}=30$ at $T = 0.02$ Kelvin; and (c) The single ring (solid line) and the double ring (dashed line) currents for $N_{s}=30$ at $T = 20.0$ Kelvin.}
\end{figure}

\begin{figure}
\includegraphics[width=4in]{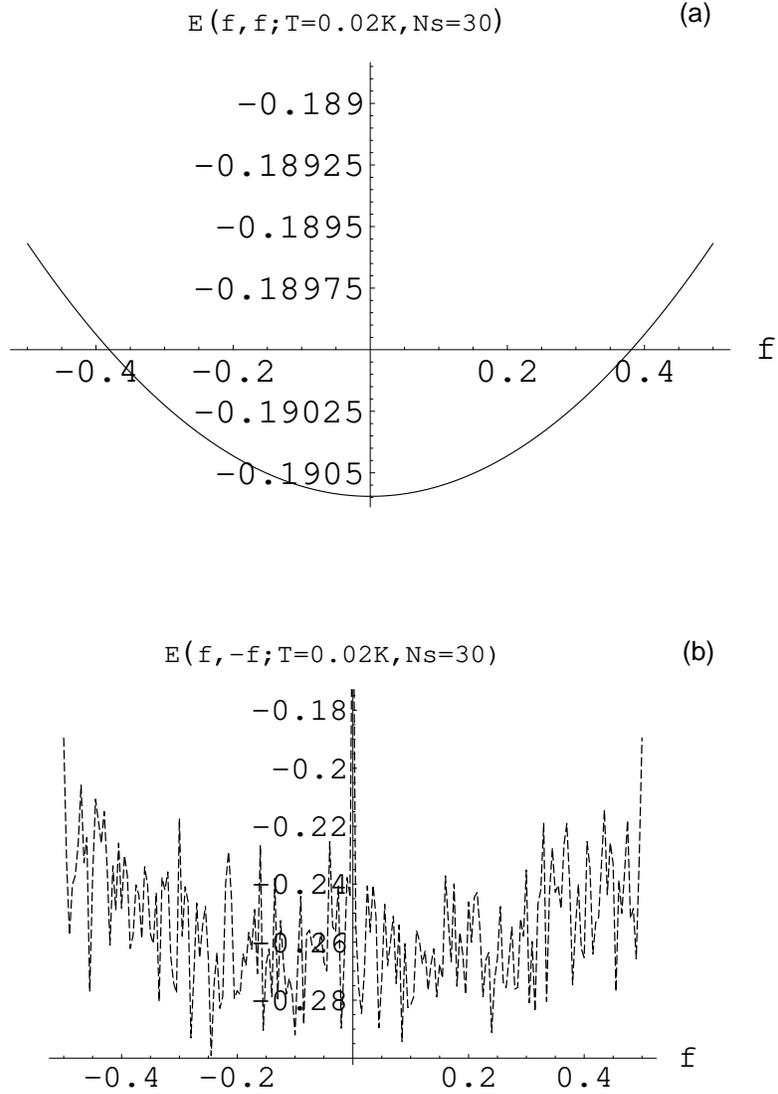}
\caption{(a) The total energy for $equal$ fluxes, $f=\hat{\varphi_1}=\hat{\varphi_2}$ $E^{(g=2)}(f,f;N_{s}=30,T=0.02 K)$; and (b) The total energy for $opposite$ fluxes, $f=-\hat{\varphi_1}=\hat{\varphi_2}$ for $30$ sites at $T=0.02$ Kelvin  $E^{(g=2)}(-f,f;N_{s}=30,T=0.02 K)$. }
\end{figure}

\end{document}